\documentstyle[12pt,epsfig]{article}
\columnwidth\textwidth

\begin{document}
\renewcommand{\theequation}{\arabic{equation}}
\parskip=4pt plus 1pt
\newcommand{\be}{\begin{equation}}
\newcommand{\ee}{\end{equation}}
\newcommand{\bea}{\begin{eqnarray}}
\newcommand{\eea}{\end{eqnarray}}
\newcommand{\co}{\; \; ,}
\newcommand{\nn}{\nonumber \\}
\newcommand{\pnn}{\nonumber}
\newcommand{\scs}{\co \;}
\newcommand{\per}{ \; .}
\newcommand{\unith}{{\bf{\mbox{1}}}}
\newcommand{\Pc}{\bar{\phi}}
\newcommand{\nonoverline}{}
\newcommand{\mput}[1]{\mbox{\tiny{#1}}}
\newcommand{\beq}{\begin{equation}}
\newcommand{\eeq}{\end{equation}}
\newcommand{\beqa}{\begin{eqnarray}}
\newcommand{\eeqa}{\end{eqnarray}}
\newcommand{\beqan}{\begin{eqnarray*}}
\newcommand{\eeqan}{\end{eqnarray*}}
\newcommand{\ba}{\begin{array}}
\newcommand{\ea}{\end{array}}
\newcommand{\ul}{\underline}
\newcommand{\ol}{\overline}
\newcommand{\olc}{\bar}
\newcommand{\Ra}{\Rightarrow}
\newcommand{\ve}{\varepsilon}
\newcommand{\vp}{\varphi}
\newcommand{\wt}{\widetilde}
\newcommand{\wh}{\widehat}
\newcommand{\cL}{{\cal L}}
\newcommand{\dfrac}{\displaystyle \frac}
\newcommand{\grts}{\stackrel{>}{_\sim}}
\newcommand{\lets}{\stackrel{<}{_\sim}}
\newcommand{\del}{\partial}
\newcommand{\dis}{\displaystyle}
\newcommand{\bla}{\left\langle}
\newcommand{\bra}{\right\rangle}
\newcommand{\no}{\nonumber}
\newcommand{\bdm}{\begin{displaymath}}
\newcommand{\edm}{\end{displaymath}}
\newcommand{\lgl}{\langle}
\newcommand{\rgl}{\rangle}

\thispagestyle{empty}
\begin{flushright}
LU TP 98-17\\
UWThPh-1998-48\\
hep-ph/9808421\\
August 1998
\end{flushright}
\vspace{2cm}
\begin{center}
\begin{Large}
Double Chiral Logs*
 \\[1cm]
\end{Large}
J. Bijnens$^1$, G. Colangelo$^{2\; \#}$ and G. Ecker$^3$ \\[2cm]
${}^1$ Dept. of Theor. Phys., Univ. Lund, S\"olvegatan 14A, S--22362 Lund, Sweden\\
${}^2$ INFN -- Laboratori Nazionali di Frascati, P.O. Box 13, I--00044
Frascati, Italy \\
${}^3$ Inst. Theor. Phys., Univ. Wien, Boltzmanng. 5, A--1090 Wien,
Austria\\[1cm]

\end{center}

\begin{abstract}
\noindent
We determine the full structure of the leading (double-pole) divergences
of $O(p^6)$ in the meson sector of chiral perturbation theory. The
field theoretic basis for this calculation is described.
We then use an extension of this result to determine the
$p^6$ contributions containing
double chiral logarithms ($L^2$), single logarithms times
$p^4$ constants ($L\times L_i^r$) and products of two $p^4$ constants
$(L_i^r\times L_j^r)$ for $F_\pi$, $F_K/F_\pi$, $K_{l3}$ and $K_{e4}$ 
form factors. Numerical results are presented for these quantities.
\end{abstract}
\setcounter{page}{0}

\vfill
\noindent * Work supported in part by TMR, EC-Contract No. 
ERBFMRX-CT980169 \\(EURODA$\Phi$NE).\\
\noindent $^\#$ 
Address after september 1 1998: Institut f\"ur Theoretische Physik der
Universit\"at Z\"urich, Winterthurerstr. 190, CH--8057 Z\"urich--Irchel.\\
\clearpage
\noindent
\paragraph{1.}  Chiral logs arise in the process of
renormalization. For instance, in dimensional regularization a
renormalization scale $\mu$ is introduced to ensure the 
correct dimension of amplitudes in $d$ dimensions. 
One-loop divergences then always appear in the scale independent 
combination
\begin{equation}
{\cal X} =
\displaystyle\frac{\mu^{d-4}}{(4\pi)^2}\left[{1\over d-4}+{1\over 2} 
\ln{M^2\over \mu^2}+\dots \right]
\label{eq:spole}
\end{equation} 
with a characteristic meson mass $M$. 
Renormalization consists in canceling the pole in 
$d-4$ by the divergent part of the tree-level amplitude of $O(p^4)$ 
and replacing it by a combination of scale dependent low-energy 
constants from $\cL_4$, the next-to-leading term in the low-energy
expansion of the effective chiral Lagrangian $\cL_{\rm eff}=\cL_2 +
\cL_4 + \cL_6 + \dots$. In many cases, the chiral logs 
$\ln{M^2/\mu^2}$ make sizable contributions 
for a typical scale $\mu$ of $O(M_\rho)$.

At $O(p^6)$, the leading divergences are double poles
accompanied by double chiral logs, the leading infrared
singularities of $O(p^6)$. The double logs are again numerically
important in general, e.g., for $S$-wave threshold parameters 
in $\pi\pi$ scattering \cite{col95,bcegs97}. As a by-product
of the complete renormalization of the generating functional of $O(p^6)$ 
\cite{bce2}, we present here the double chiral logs in full
generality for chiral $SU(n)$. As will be shown below, the double logs
($L^2$) come together with terms of the form $L\times L_i^r$ and
products  $L_i^r\times L_j^r$ where the $L_i^r$ are the renormalized
low-energy constants of $O(p^4)$ (we use the $SU(3)$
notation here for simplicity). It is then very natural to include
such terms in the numerical analysis especially because they are often
comparable to or even bigger than the proper double-log terms.

The purpose of this letter is twofold. First, our general double-pole
divergence Lagrangian for $SU(n)$ may serve as a check of existing and
yet to be performed two-loop calculations of $O(p^6)$. Secondly,
the numerical analysis may provide hints as to where large $p^6$
corrections are to be expected. Of course, the partial $p^6$ results
presented here are not a substitute for the full
expressions of $O(p^6)$. But once the double-pole divergence Lagrangian
has been determined from a one-loop calculation (see below), the
numerical applications come at almost no cost compared to the full
$p^6$ calculations. 

\begin{figure}
\centerline{\epsfig{file=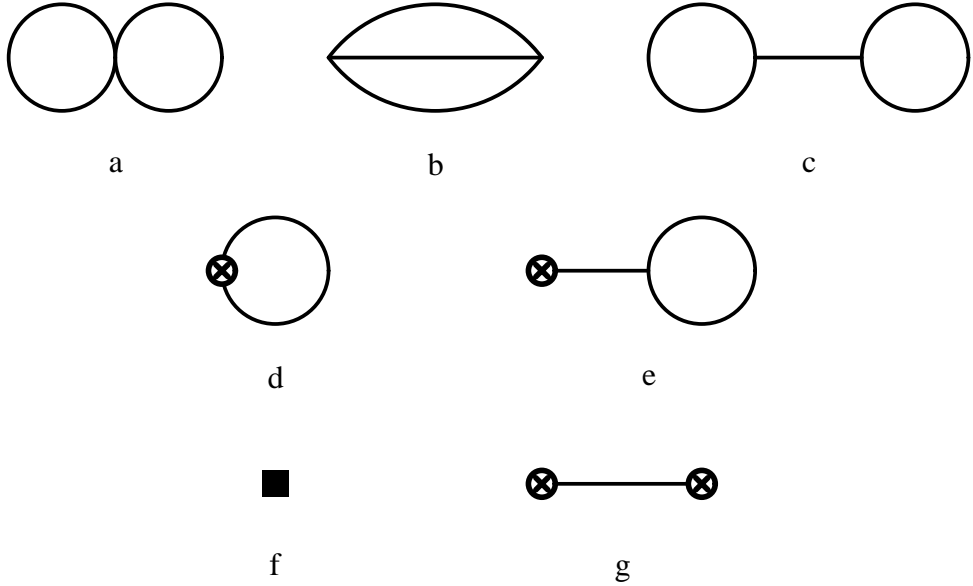,height=8cm}}
\caption{Diagrams contributing to the generating functional of $O(p^6)$.
The propagators and vertices carry the full tree structure associated
with the lowest--order Lagrangian $\cL_2$. Normal vertices are from
$\cL_2$, crossed circles denote vertices from $\cL_4$ and the square in diagram
f stands for a vertex from $\cL_6$.}\label{fig:p6}
\end{figure}

\noindent
\paragraph{2.}
In a mass independent regularization scheme like dimensional
regularization, the divergent parts are polynomials in masses and 
external momenta. In the present case, the coefficients 
of those polynomials are renormalized by the general chiral Lagrangian
of $O(p^6)$. Therefore, the divergences themselves can be cast into the 
form of $\cL_6$ with divergent coefficients that receive  
contributions from the diagrams shown in Fig.~\ref{fig:p6}.

There are two types of diagrams in Fig.~\ref{fig:p6}, reducible
and irreducible ones, both divergent in general. However, as
will be shown below, one can always choose $\cL_4$ in such a way as
to make the sum of the reducible diagrams c,e,g finite. Even so, 
these diagrams give rise to double logs in general.
These double logs are
completely determined by the tree-level contribution g proportional to
products of low-energy constants of $O(p^4)$. 

The tree-level diagram f being trivial, we are left with the irreducible
loop diagrams a,b,d in Fig.~\ref{fig:p6}. Starting with the
two-loop diagrams a, b, the corresponding divergences in
a given coefficient of $\cL_6$ are of the general form
\begin{eqnarray} 
& & x(d) \displaystyle\frac{\mu^{2(d-4)}}{(4\pi)^4}\left[{1\over d-4}
+{1\over 2} \ln{M^2\over \mu^2}+\dots \right]^2 \nn
&=& x(d) \displaystyle\frac{\mu^{2(d-4)}}{(4\pi)^4}\left[{1\over (d-4)^2}+
\displaystyle\frac{\ln{M^2/\mu^2}}{d-4}  +
{1\over 4} (\ln{M^2\over \mu^2})^2+\dots \right]
\label{eq:2loopd}
\end{eqnarray} 
where $x(d)$ is a function of $d$ that depends on the specific 
coefficient under consideration.

The problematic piece in (\ref{eq:2loopd}) is the nonlocal divergence
${x_0\ln{M^2/\mu^2}}/({(4\pi)^4 (d-4)})$
that cannot be renormalized with a local action of $O(p^6)$.
Here, $x_0$ is the leading coefficient in the
Taylor expansion of $x(d)$ around $d=4$:
$x(d)=x_0+x_1(d-4)+\cdots~.$
Therefore, in accordance with general theorems of
renormalization theory \cite{collins}, 
the nonlocal divergence 
must cancel with other 
contributions. Since the sum of
reducible diagrams in Fig.~\ref{fig:p6} is finite with our choice of 
$\cL_4$, the cancellation can only come from diagram d that is
proportional to the low-energy constants $L_i$ . These constants are themselves
divergent (to renormalize the one-loop divergences). In 
$d$ dimensions, they can be written
\begin{equation}
L_i(d)=\mu^{d-4}\left[\displaystyle\frac{\Gamma_i}{(4\pi)^2(d-4)}+
L_i^r(\mu)+\dots  \right]
\label{eq:Lid}
\end{equation} 
with known coefficients $\Gamma_i$ and renormalized constants
$L_i^r(\mu)$ \cite{gl85}. The divergences due to the one-loop diagram d 
in Fig.~\ref{fig:p6} are then given by the product of (\ref{eq:spole}) 
and (\ref{eq:Lid}) with appropriate coefficients $y_i(d)$:
\bea
-{y_i(d)\over 2}
{\cal X} L_i(d)& =&
-{y_i(d)\over 2} \displaystyle\frac{\mu^{2(d-4)}}{(4\pi)^4}
\left[{\Gamma_i\over(d-4)^2}+\displaystyle\frac{\Gamma_i\ln{M^2/\mu^2}}
{2(d-4)} +  \displaystyle\frac{(4\pi)^2 L_i^r(\mu)}{d-4}\right.\nn
\label{eq:1loopd} 
&+& \left.{(4\pi)^2\over 2} L_i^r(\mu)\ln{M^2/\mu^2} +\dots \right]~.
\eea
Summation over $i$ is implied and we have chosen the numerical 
factor $-1/2$ in (\ref{eq:1loopd}) to conform to the notation of 
Ref.~\cite{bcegs97}. Cancellation of the nonlocal divergences in 
(\ref{eq:2loopd}) and (\ref{eq:1loopd}) is guaranteed by Weinberg's 
conditions \cite{wein79}
\begin{equation} 
x_0={1\over4}y_{i0}\Gamma_i ~,
\label{eq:weinrel}
\end{equation} 
with $y_{i0}$ the leading coefficient in the Taylor expansion of
$y_i(d) = y_{i0} + (d-4)y_{i1}+\ldots$.
In the chiral  Lagrangian
of $O(p^6)$ for $SU(n)$, there are 115 independent terms 
(including three contact terms) \cite{bce2}. We have verified 
the corresponding 115 relations (\ref{eq:weinrel}) by explicit 
calculation of both two- and one-loop divergences.

The divergences in the sum of (\ref{eq:2loopd}) and (\ref{eq:1loopd}) 
now have the proper local form to be renormalized by the counterterm
action of $O(p^6)$ represented by diagram f in Fig.~\ref{fig:p6}. 
After renormalization, we are left with finite parts 
\begin{equation} 
\displaystyle\frac{x_0}{4(4\pi)^4}(\ln{M^2\over \mu^2})^2-
\displaystyle\frac{y_{i0}L_i^r(\mu)}{4(4\pi)^2}\ln{M^2\over \mu^2}
+\dots
\label{eq:finparts}
\end{equation} 
The terms listed explicitly are the double-log contribution and
all products of renormalized low-energy constants with a chiral log. 
With the help of (\ref{eq:weinrel}), these two terms can be combined 
to give
\begin{equation} 
\displaystyle\frac{y_{i0}}{16}\left[\Gamma_i L^2 - 4 L_i^r(\mu)
L\right] 
:= -\frac{y_{i0}}{16}\,k_i
\qquad , \qquad L={1\over (4\pi)^2}\ln{M^2\over \mu^2}~.
\label{eq:ki}
\end{equation}
The combinations 
$k_i$ \cite{col95,bcegs97}
always appear together in the renormalized coefficients. They are
process independent generalizations of the double logs and their 
coefficients are completely calculable from the one-loop
diagram d in Fig.~\ref{fig:p6}. 
The main reason for using the $k_i$ together with the products
$L_i^r\times L_j^r$ in numerical applications instead of only the 
double logs is a practical one. Very often, in particular for chiral
$SU(3)$ for the conventional choices of $M/\mu$, the additional terms 
are numerically at least as important as the double logs
themselves. 

Before discussing the actual calculation in more detail, we present
the final result below. Since up to 11 of the $k_i$ can appear in each 
of the 115 coefficients of $\cL_6$, the complete Lagrangian as a 
function of the $k_i$ cannot be reproduced here. Instead, we
only write down the proper double-log Lagrangian for chiral $SU(n)$
in Minkowski space, 
i.e. dropping the terms\footnote{The full expression with all $k_i$
and all terms can be obtained from the authors on request.}
of the form $ L_i^r(\mu)L$. We also drop all terms that do not 
contribute to processes with up to four mesons or one external 
field and three mesons.
\begin{eqnarray} 
\lefteqn{\cL_{\rm double-log}\, =\, \displaystyle\frac{L^2}{F^2}\{
 -(7/64 + 5/1152~n^2 ) \lgl A h_{\mu \nu} h^{\mu \nu} \rgl  
       - 41/2304~n ~\lgl A \rgl \lgl h_{\mu \nu} h^{\mu \nu} \rgl}
\nn&&  
       + 3/64 \, \lgl h_{\mu \nu} u_\rho h^{\mu \nu} u^\rho \rgl
 +1/288\,n  \,\lgl h_{\mu \nu} u_\rho \rgl \lgl h^{\mu \nu} u^\rho \rgl 
       - 1/16 \,\lgl h_{\mu \nu} \left(u_\rho h^{\mu \rho}
           u^\nu + u^\nu h^{\mu\rho} u_\rho \right) \rgl \nn 
&& - 1/384\,n \,\lgl h_{\mu \nu} u_\rho \rgl \lgl h^{\mu \rho} u^\nu \rgl  
+ 1/144\,n^2 \,\lgl A^2 \chi_+ \rgl  
+ 1/18\,n  \, \lgl A^2 \rgl \lgl \chi_+ \rgl
\nn&&
 - 1/64\,n  \,\lgl A \rgl \lgl A \chi_+ \rgl  
- 1/128 \,\lgl A \rgl^2 \lgl \chi_+ \rgl  
- ( 3/32 - 17/1152\,n^2 )\lgl A u_\mu \chi_+ u^\mu \rgl  \nn
&& + ( 3/32 - 1/72\,n^2 ) \lgl \chi_+ u_\mu u_\nu u^\mu u^\nu \rgl  
- 1/18\,n  \,\lgl \chi_+ \rgl \lgl  u_\mu u_\nu u^\mu u^\nu \rgl \nn
&& + 1/32\,n \,\lgl \chi_+ u_\mu u_\nu \rgl \lgl u^\mu u^\nu \rgl  
+ 1/64 \,\lgl \chi_+ \rgl \lgl u_\mu u_\nu \rgl^2  
-  19/2304\,n^2 \,\lgl \chi_+ h_{\mu \nu} h^{\mu \nu} \rgl  
\nn&&
-  67/2304\,n \,\lgl \chi_+ \rgl \lgl h_{\mu \nu} h^{\mu \nu} \rgl  
+  3/128 \,\lgl A \chi_+^2 \rgl  
- 3/64\,n^{-1} \,\lgl A \chi_+ \rgl \lgl \chi_+ \rgl  
\nn&&
- (3/128\,n^{-1} + 1/128\,n )  \lgl A \rgl \lgl \chi_+^2 \rgl 
  - (1/256 - 1/32\,n^{-2} ) \lgl A \rgl \lgl \chi_+ \rgl^2 
\nn&&
- (3/64 - 1/256\,n^2 )\lgl \chi_+ u_\mu \chi_+ u^\mu \rgl  
 + ( 5/64\,n^{-1} - 1/256\,n )\lgl \chi_+ u_\mu \rgl^2  
+ 1/8\,n^{-2} \,\lgl \chi_+^3 \rgl  \nn 
&&- ( 1/8\,n^{-3} + 1/16\,n^{-1} )\lgl \chi_+^2 \rgl \lgl \chi_+ \rgl  
+ (1/32\,n^{-4} + 1/32\,n^{-2} ) \lgl \chi_+ \rgl^3 \nn
&&- ( 1/192 - 1/768\,n^2 )i \lgl \chi_- \{ h_{\mu \nu},u^\mu u^\nu\} \rgl
+ 1/384\,n  \,i \lgl \chi_-  h_{\mu \nu} \rgl \lgl u^\mu u^\nu \rgl  \nn
&&- ( 1/8\,n^{-1} + 7/384\,n ) i \lgl h_{\mu \nu} u^\mu u^\nu
           \rgl \lgl \chi_- \rgl  
+ 13/96 \,i \lgl h_{\mu \nu} u^\mu \chi_- u^\nu \rgl  
\nn&&
+ 5/128\,n \,i \lgl h_{\mu \nu} u^\mu \rgl \lgl \chi_- u^\nu \rgl  
- (5/48 - 1/32\,n^{-2} - 1/1152\,n^2 ) \lgl A \chi_-^2 \rgl\nn  
&&+ ( 3/32\,n^{-1} + 19/1152\,n )\lgl A \chi_- \rgl \lgl \chi_- \rgl
- ( 1/64\,n^{-1} + 1/72\,n )\lgl A \rgl \lgl \chi_-^2 \rgl  \nn  
&&        + ( 31/2304 - 1/16\,n^{-2} )\lgl A \rgl \lgl \chi_- \rgl^2  
       + ( 5/192 - 1/32\,n^{-2} - 1/768\,n^2 ) \lgl u_\mu \chi_- u^\mu
           \chi_- \rgl  \nn
&&       + ( 1/32\,n^{-1} - 5/1152\,n ) \lgl u_\mu \chi_- \rgl^2  
       - ( 1/32 - 5/32\,n^{-2} + 5/1152\,n^2 ) \lgl \chi_-^2 \chi_+ \rgl  \nn
&&   - ( 1/16\,n^{-3}\! + 29/1152\,n ) \lgl \chi_+ \rgl \lgl \chi_-^2 \rgl  
      - ( 1/16\,n^{-3}\! + 1/16\,n^{-1}\! - 7/288\,n )\lgl \chi_+ \chi_-
           \rgl \lgl \chi_- \rgl  \nn
&&    + ( 1/48 + 1/32\,n^{-4} + 1/64\,n^{-2} ) \lgl \chi_+ \rgl
           \lgl \chi_- \rgl^2  
  + ( 1/64 + 1/192\,n^2 )i \lgl \chi_- \{ \nabla_\mu \chi_+,u^\mu \} \rgl  \nn
&&       + ( 1/32\,n^{-1} - 5/192\,n )i \lgl \chi_- \rgl \lgl \nabla_\mu
           \chi_+ u^\mu \rgl  
       - (1/32\,n^{-1} - 11/192\,n )i \lgl \nabla_\mu \chi_+ \rgl
           \lgl \chi_- u^\mu \rgl  \nn
&&       - (1/32 - 1/128\,n^2 )\lgl \nabla_\mu \chi_+ \nabla^\mu \chi_+ \rgl  
       +  3/128\,n \,\lgl \nabla_\mu \chi_+ \rgl^2  
\nn&&
    + 1/144\,n^2 \,i \lgl f_{+ \mu \nu} \{\chi_+, u^\mu u^\nu \} \rgl  
       + 1/9\,n \,i \lgl \chi_+ \rgl \lgl f_{+ \mu \nu} u^\mu u^\nu \rgl  
       + 1/72\,n^2 \,i \lgl f_{+ \mu \nu} u^\mu \chi_+ u^\nu \rgl  \nn
&&
       - (1/48 - 5/1152\,n^2 ) \lgl f_{- \mu \nu} \left(h^{\nu
           \rho} u_\rho u^\mu + u^\mu u_\rho h^{\nu \rho} \right) \rgl  
    - 1/576\,n \,\lgl f_{- \mu \nu} h^{\nu \rho} \rgl \lgl u^\mu u_\rho \rgl  
\nn&&
   + 17/288\,n \,\lgl f_{- \mu \nu} u^\mu \rgl \lgl h^{\nu \rho} u_\rho \rgl
    + ( 1/48 + 1/192\,n^2 ) \lgl f_{- \mu \nu} \left(u^\mu h^{\nu
           \rho} u_\rho +  u_\rho h^{\nu \rho} u^\mu \right) \rgl  
  \nn
&& - ( 1/16 + 1/2304\,n^2 ) i \lgl f_{- \mu \nu} [\chi_-, u^\mu u^\nu] \rgl  
   - 35/1152\,n \,i \lgl f_{- \mu \nu} u^\nu \rgl \lgl u^\mu \chi_- \rgl  \nn
&&  - 5/288\,n^2 \,\lgl f_{- \mu \nu} \{ \nabla^\mu \chi_+, u^\nu \} \rgl  
     - 31/288\,n \,\lgl \nabla^\mu \chi_+ \rgl \lgl f_{- \mu \nu} u^\nu \rgl \nn
&& 
    - (1/96 + 1/288\,n^2 ) i \lgl \nabla_\rho f_{+ \mu \nu} 
           [h^{\mu \rho}, u^\nu ] \rgl
    - (1/96 + 1/384\,n^2 ) i \lgl \nabla^\mu f_{+ \mu \nu} 
           [h^{\nu \rho}, u_\rho ] \rgl \}
\label{eq:Ldlog}
\end{eqnarray} 
with $A=u_\mu u^\mu$ and $h_{\mu\nu}=\nabla_\mu u_\nu + 
\nabla_\nu u_\mu$. $F$ is the meson decay constant in the chiral limit
and the quantities $u_\mu$, $\nabla_\mu$, $\chi_\pm$,
$f_\pm^{\mu\nu}$ are defined as usual (e.g., in Ref.~\cite{egpr89}).

In the chosen basis for the $SU(n)$ Lagrangian of $O(p^6)$, 109 of 
altogether 115 coefficients
are divergent. For $n=2$ or $3$ \cite{fs96}, there are of course fewer
independent terms but we postpone the explicit discussion of
$\cL_6$ for $n=2,3$ to a separate publication \cite{bce2}.

\noindent
\paragraph{3.} In this section, we sketch the main steps for
extracting the double logs from the generating functional of $O(p^6)$.
All technical details will be deferred to a separate publication
\cite{bce2}. 

In its most general form, the effective Lagrangian of $O(p^4)$ for 
chiral $SU(n)$ is given by
\beqa
{\cal L}_4 & = & C_0 \lgl u_\mu u_\nu u^\mu u^\nu \rgl 
        + C_1 \lgl u_\mu u^\mu \rgl^2 
        + C_2 \lgl u_\mu u_\nu\rgl\lgl u^\mu u^\nu \rgl 
        + C_3 \lgl u_\mu u^\mu u_\nu u^\nu \rgl \nn
& &     + C_4 \lgl u_\mu u^\mu \rgl\lgl \chi_+ \rgl
        + C_5 \lgl u_\mu u^\mu \chi_+ \rgl 
        + C_6 \lgl \chi_+ \rgl^2
        + C_7 \lgl \chi_- \rgl^2
        + {1\over 2} C_8 \lgl \chi_+^2 + \chi_-^2 \rgl\nn
& &     - i C_9 \lgl f_+^{\mu\nu}u_\mu u_\nu \rgl
        + {1\over 4} C_{10}\lgl f_+^{\mu\nu}f_{+\mu\nu} - 
f_-^{\mu\nu}f_{-\mu\nu}\rgl\nn
& &     +i C_{11} \lgl \hat\chi_- (\nabla^\mu u_\mu - {i\over
2}\hat\chi_-)\rgl 
        + C_{12}  \lgl(\nabla^\mu u_\mu - {i\over
2}\hat\chi_-)^2\rgl
        + \mbox{~contact~terms} 
\label{eq:L4}
\end{eqnarray}
with $\hat\chi_- := \chi_- - \lgl\chi_- \rgl/n$. The terms
with coefficients $C_{11}, C_{12}$ vanish at the classical solution
defined by the equation of motion (EOM)
$
\nabla^\mu u_\mu - i \hat\chi_-/2=0 ~.
$
All chiral Lagrangians of $O(p^4)$ can be
written in the form (\ref{eq:L4}) but there is of course some
redundancy for $n=2$ or $3$. 
The Laurent expansion analogous to (\ref{eq:Lid}) is
\begin{equation}
C_i(d)=\mu^{d-4}\left[\displaystyle\frac{\Sigma_i}{(4\pi)^2(d-4)}+
C_i^r(\mu)+\dots  \right]
\label{eq:Cid}
\end{equation} 
with known coefficients $\Sigma_i$ \cite{gl85}.

The calculation of both reducible and one-loop diagrams in 
Fig.~\ref{fig:p6} proceeds along standard lines by expanding 
the chiral actions of $O(p^2)$ and $O(p^4)$ around the classical 
solution defined by the equation of motion. 
In particular, we
need the expansion of $S_4$ up to second order in the fluctuation
variables $\xi$ defined in the usual way \cite{gl85}:
\begin{equation}
S_4[\phi]=S_4[\phi_{\rm cl}]+S_{4i}[\phi_{\rm cl}]\xi_i
+{1\over 2}S_{4ij}[\phi_{\rm cl}]\xi_i\xi_j + O(\xi^3)
\end{equation}
$$
u(\phi) = u(\phi_{\rm cl})e^{i\xi(\phi)/2}, ~~
\xi^\dagger  = \xi, ~~ \langle \xi \rangle = 0, \quad 
\xi(\phi_{\rm cl})= 0,\quad
\xi={1\over \sqrt{2}}\lambda_i \xi_i~,\quad \lgl \lambda_i \lambda_j
\rgl = 2 \delta_{ij}~,
$$
with $\lambda_i$ the generators of $SU(n)$ in the fundamental
representation. 

We will not write down the explicit expressions for $S_{4i}$ and
$S_{4ij}$ here (see Ref.~\cite{bce2}), but concentrate on the
EOM terms in (\ref{eq:L4}). As in the actual calculations, we switch
to Euclidean space for the rest of this section. For investigating 
the influence of EOM terms, one needs the expansion
\begin{equation}
\nabla_\mu u_\mu + {i\over 2}\hat\chi_-= {1\over \sqrt{2}}\lambda_i 
(-d_\mu d_\mu + \sigma)_{ij}\xi_j+ O(\xi^2)~.
\label{eq:EOMxi}
\end{equation}
In (\ref{eq:EOMxi}), $(-d_\mu d_\mu + \sigma)_{ij}=G_{ij}^{-1}$
is the inverse of the full propagator\footnote{This is the propagator
occurring in the functional diagrams of Fig.~\ref{fig:p6}.} of
$O(p^2)$ in the presence of external fields.

Turning first to the last term in (\ref{eq:L4}), we infer from 
(\ref{eq:EOMxi}) that $C_{12}$ cannot contribute to the reducible
diagrams e,g in Fig.~\ref{fig:p6}. Moreover, the one-loop diagram d
proportional to  $C_{12}$ vanishes in dimensional regularization
because of $\delta^d(0)=0$. In other schemes, the one-loop
contribution would be a quartic (and sixth-order) divergence that can
be absorbed in $\cL_6$. Chiral logs do not occur and
we can set $C_{12}=0$ without loss of generality.

The situation is more subtle for $C_{11}$. The explicit factor
$G_{ij}^{-1}$ in (\ref{eq:EOMxi}) implies that all
tree-level contributions from diagram g involving $C_{11}$ are local,
i.e. these terms can always be absorbed in $\cL_6$. An explicit
calculation shows in addition that diagrams d and e cancel exactly for
the vertex associated with $C_{11}$. The final result is that $C_{11}$
appears only in a (local) tree-level functional that has no
bearing on chiral logs. Since we are free to choose any
value for $C_{11}$, the choice $C_{11}=0$ is the most convenient one
for the calculation of the generating functional (for $n=2,3$, 
this corresponds to the Lagrangians of Refs.~\cite{gss88} and 
\cite{gl85}, respectively). 

Diagrams c,e,g make up the reducible part of the generating
functional of $O(p^6)$:
\begin{equation}
Z_6^{\rm red}= -{1\over F^2}\left(S_{4i}+{1\over
2}D_{ikl}^{(3)}G_{kl}\right)G_{ij}\left(S_{4j}+{1\over
2}D_{jmn}^{(3)}G_{mn}\right)
\label{eq:Zred}
\end{equation}
where $D_{ikl}^{(3)}$ represents the cubic functional vertices in
diagrams c,e (and b, for that matter). For $C_{11}=0$, the 
combination
$$
S_{4i}+{1\over 2}D_{ikl}^{(3)}G_{kl}
$$
turns out to be finite and scale independent by itself (see also
Ref.~\cite{em96}). In other
words, the one-loop divergences in c and e are cancelled by the
divergent parts of the coefficients $C_i ~(i=0,\dots,10)$ given in
(\ref{eq:Cid}).
This is precisely what one
expects: the one-loop divergences are already taken care of by the 
renormalization at $O(p^4)$. Note however that this would not be 
the case for $C_{11}\ne 0$ in general.

Due to the finiteness and $\mu$-independence of the reducible
functional (\ref{eq:Zred}), the renormalized low-energy constants
$C_i^r(\mu)$ always appear in the scale independent combination
\begin{equation} 
C_i^r(\mu) - {1\over 2}\Sigma_i L
\end{equation}
with $L$ defined in (\ref{eq:ki}). Therefore, the dependence of 
$Z_6^{\rm red}$ on chiral logs is completely fixed
by the dependence on the $C_i^r(\mu)$ via diagram g. In the actual
applications, chiral logs from this source will appear in mass
and wave function renormalization, $F_\pi$, etc.\footnote{In
general, there
can also be nonlocal double-log contributions. For instance, there
will be such contributions for processes with six external mesons.}

Finally, the one-loop functional associated with diagram d is given by
\begin{equation}
Z_6^{L=1}=\displaystyle\frac{1}{F^2}S_{4ij}G_{ij}~.
\end{equation}
Decomposing the propagator $G_{ij}(x,y)$ in the usual way \cite{jo82}
into two parts, being singular and finite in the coincidence limit
$x\to y$, respectively, $Z_6^{L=1}$ exhibits two types of divergences
as sketched in Eq.~(\ref{eq:1loopd}). The nonlocal ones cancel with
corresponding divergences from the irreducible diagrams a,b. The local
divergences (of a,b and d) are cancelled by counterterms in 
$\cL_6$. Following the arguments of the previous section,
they also determine the finite parts given in
(\ref{eq:finparts}). Keeping only the double-log contributions,
the final result is expressed in terms of the double-log Lagrangian 
(\ref{eq:Ldlog}).

\noindent
\paragraph{4.}
Let us now turn to some applications of the above results.
For the case of two flavours, $n=2$, a number of full two-loop
calculations already exists (a short review can be found in
\cite{bijnensmainz}). We have checked that the double logarithms agree with
those calculated for the following quantities:
$\pi\pi$-scattering \cite{bcegs97},
$F_\pi$ and $M_\pi^2$ \cite{bcegs97,buergi},
radiative pion decay \cite{BT} and the pion vector and scalar form factors
\cite{bct98}.

In the case of three flavours we agree with the double logarithms as calculated
for the vector two-point function \cite{gk1}, for $M_\pi^2$, $M_\eta^2$,
$F_\pi$ and $F_\eta$ \cite{gk2}. To emphasize that these agreements
provide nontrivial checks on our calculations, we show here
the $p^6$ contributions from the $k_i$ defined in (\ref{eq:ki}) and from 
terms of the type $L_i^r L_j^r$ for $F_\pi/F$:
\bea
\lefteqn{F_\pi^4(F_\pi/F\,)^{(6)}\,=\,
        M_\pi^4   ( 56 L_4^{r2} + 112 L_4^r L_5^r - 64 L_4^r L_6^r 
 - 64 L_4^r L_8^r + 56  L_5^{r2} - 64 L_5^r L_6^r}&&
\nn&&
 - 64 L_5^r L_8^r
 + 65/9\,k_1 + 73/18\,k_2 + 32/9\,k_3 - 20/
         3\,k_4 - 5\,k_5 )
       + M_\pi^2 M_K^2   ( 224 L_4^{r2}
\nn&&
 + 160 L_4^r L_5^r - 256 L_4^r L_6^r - 128 L_5^r L_6^r
          - 16/9\,k_1 - 4/9\,k_2 - 4/9\,k_3 - 37/6\,k_4
 - 5/2\,k_5 )
\nn&&
       + M_K^4   ( 224 L_4^{r2} + 64 L_4^r L_5^r - 256 L_4^r L_6^r - 128 L_4^r L_8^r + 
         104/9\,k_1
 + 26/9\,k_2
\nn&&
 + 61/18\,k_3 - 29/3\,k_4 )~.
\eea
Here and in the following, as indicated by the superscript $(6)$,
we only display the partial results of $O(p^6)$ (in the sense
explained above) for all quantities considered.
In addition we always use isospin symmetry.

The result for $F_K/F_\pi$ is new:
\bea
\lefteqn{F_\pi^4(F_K/F_\pi)^{(6)}\, =\,
        M_\pi^2(M_K^2-M_\pi^2)   (   32 L_4^r L_5^r + 40 L_5^{r2}
    - 64 L_5^r L_6^r - 64 L_5^r L_8^r + 
         k_1 }
\nn&&
+ 5/2\,k_2
+ 19/12\,k_3
 - 37/12\,k_5 + 4\,k_7 + 2\,k_8 )
       + M_K^2 (M_K^2-M_\pi^2)   ( 64 L_4^r L_5^r 
+ 24 L_5^{r2}
\nn&&
 - 128 L_5^r L_6^r - 64 L_5^r L_8^r + k_1
          + 5/2\,k_2 + 23/12\,k_3 - 17/12\,k_5 - 4\,k_7 - 2\,k_8 )~.
\eea
We defer a numerical discussion to the next section.
The $p^4$ results for $F_\pi/F$ and $F_K/F_\pi$ can be found in \cite{gl85}.

We have also worked out the partial $p^6$ results for several
relevant semileptonic kaon decays. For the two $K_{l3}$ form factors
$f_+$ and $f_0$ we obtain:
\bea
\lefteqn{F_\pi^4 (f_+(t))^{(6)}\,=\,
        (M_K^2-M_\pi^2)^2    (   8  L_5^{r2} + 1/3 \,k_1 + 1/6 
         \,k_2 + 1/36 \,k_3 - \,k_4 )}&&
\nn&&
       + t M_\pi^2    ( 8  L_5^r L_9^r - \,k_1 + 1/2 \,k_2
 - 1/4\,k_3 - 
         1/3\,k_4 - 1/2\,k_5 - \,k_9)\nn&&
+ t M_K^2 ( - 8 L_5^r L_9^r -\,k_1 + 1/2 
        \,k_2 - 5/4\,k_3 - 1/3\,k_4 - 1/2\,k_9 )
\nn&&
       + t^2   ( 1/3\,k_1 - 1/6\,k_2 + 1/4\,k_3 - 1/8\,k_9 )~,
\eea
\bea
\lefteqn{F_\pi^4 (f_0(t))^{(6)}\,=\,
        (M_K^2-M_\pi^2)^2 ( 8  L_5^{r2} + 1/3 \,k_1 + 1/6 
         \,k_2 + 1/36 \,k_3 - \,k_4 )}&&
\nn&&
       + t M_\pi^2  ( 32  L_4^r L_5^r - 64  L_5^r L_6^r - 64  L_5^r
L_8^r + 48  L_5^{r2}  + 4/3 \,k_1 + 11/3 \,k_2 \nn&&
 + 22/9\,k_3 + \,k_4 - 23/6\,k_5 + 4\,k_7 + 2\, k_8)
\nn&&
+ t M_K^2 (64  L_4^r L_5^r - 128  L_5^r L_6^r - 64  L_5^r
L_8^r + 16  L_5^{r2}  + 4/3 \,k_1 + 11/3 \,k_2\nn&& 
 + 25/9\,k_3 + \,k_4 - 2/3\,k_5 - 4\,k_7 - 2\, k_8)
\nn&&
   + t^2   ( -2/3\,k_1 - 4/3\,k_2 - 8/9\,k_3 - 3/4\,k_5 )~.
\eea
The Ademollo-Gatto theorem \cite{ag64} is the reason for
the appearance of the factor $(M_K^2-M_\pi^2)^2$.
The definition of the two form factors and a discussion of experimental results
can be found in \cite{daphne}. The $p^4$ results can also be found there
and were first obtained in \cite{gl85b}.

We also derived the corresponding results for the $K_{e4}$ form factors
$F$ and $G$. The definition of all quantities appearing
here can be found in \cite{daphne,bcgke4} and the $p^4$ results
using the same notation can also be found there. They were first derived
in \cite{bijnenske4,rigke4}. 
\bea
\lefteqn{(\sqrt{2} F_\pi/M_K) F_\pi^4\, F(q^2,\nu,s_l)^{(6)}\,=\, 
  M_\pi^4 q^4   (  - 404 \,k_1 - 178 \,k_2 - {431/3} \,k_3)}&&
\nn&&
       + M_\pi^4 q^2   ( 512 L_1^r L_5^r + 128 L_2^r L_5^r + 160 L_3^r 
         L_5^r - {1904/3} \,k_1 - {2686/9} \,k_2 - {11987/54} \,k_3
\nn&&         - 80 \,k_4
          - 27 \,k_5 )
       + M_\pi^4   ( 256 L_1^r L_5^r + 128 L_2^r L_5^r + 96 L_3^r L_5^r + 
         256 L_4^{r2} + 416 L_4^r L_5^r
\nn&&
 - 512 L_4^r L_6^r - 512 L_4^r L_8^r + 56 
         L_5^{r2} - 64 L_5^r L_6^r - 64 L_5^r L_8^r - 1969/9 \,k_1
\nn&&
 - 2509/18 
         \,k_2 - 9133/108 \,k_3 - 854/9 \,k_4 - 271/12 \,k_5 - 16 \,k_6 - 
         20 \,k_7 - 18 \,k_8 )
\nn&&
       + M_\pi^2 M_K^2 q^2   (  - 512 L_1^r L_5^r - 128 L_2^r L_5^r - 160 
         L_3^r L_5^r + 437/3 \,k_1 + 229/6 \,k_2
\nn&&
 + 691/12 \,k_3 - 52 \,k_4 )
       + M_\pi^2 M_K^2   (  - 256 L_1^r L_5^r - 96 L_2^r L_5^r - 88 L_3^r L_5^r
        + 512 L_4^{r2}
\nn&&
 - 64 L_4^r L_5^r - 1024 L_4^r L_6^r - 32 L_5^{r2} - 128 
         L_5^r L_6^r + 1000/9 \,k_1 + 247/18 \,k_2
\nn&&
 + 9005/216 \,k_3 - 100/
         9 \,k_4 + 73/12 \,k_5 - 44 \,k_6 + 20 \,k_7 - 12 \,k_8 )
\nn&&
       + M_\pi^2 s_l q^2   ( \,k_1 + 71/6 \,k_2 - 5/12 \,k_3 - 9/2 
         \,k_9 )
      + M_\pi^2 s_l   (  - 32 L_2^r L_5^r - 8 L_3^r L_5^r
\nn&&
  + 8 L_5^r L_9^r + 4/
         3 \,k_1 + 337/18 \,k_2 + 319/216 \,k_3 + 3/2 \,k_5 - 43/8 
         \,k_9 )
\nn&&       + M_\pi^2 \nu q^2   (  - 5 \,k_1 - 15/2 \,k_2 - 19/4 \,k_3 )
       + M_\pi^2 \nu   (  - 8 L_3^r L_5^r - 9/2 \,k_1 - 33/4 \,k_2
\nn&& - 19
         /6 \,k_3 - 2 \,k_4 )
       + M_K^4   (  - 32 L_2^r L_5^r - 8 L_3^r L_5^r + 8 L_5^{r2} - 1/
         3 \,k_1 - 6 \,k_2 - 61/72 \,k_3 
\nn&&
- 4/3 \,k_4 )
       + M_K^2 s_l   ( 32 L_2^r L_5^r + 8 L_3^r L_5^r - 8 L_5^r L_9^r + 41/6
          \,k_2 - 11/24 \,k_3 - 7/8 \,k_9 )
\nn&&
       + M_K^2 \nu   ( 8 L_3^r L_5^r - 19/6 \,k_1 - 17/6 \,k_2 - 145/
         72 \,k_3 - \,k_4 )
       + s_l^2   (  - 1/3 \,k_2 + 1/3 \,k_3
\nn&&
 + 1/4 \,k_9 )
       + s_l \nu   ( \,k_1 + 7/4 \,k_2 + 9/8 \,k_3 )
       + \nu^2   (  - 5/12 \,k_2 + 7/24 \,k_3 )~,
\eea
\bea
\lefteqn{(\sqrt{2} F_\pi/M_K)F_\pi^4\, G(q^2,\nu,s_l)^{(6)}\,=\, 
        M_\pi^4 q^4   (  - 20 \,k_2 - 2 \,k_3 )}&&
\nn&&
       + M_\pi^4 q^2   (  - 32 L_3^r L_5^r - 22/3 \,k_1 - 119/3 
         \,k_2 - 34/9 \,k_3 - 8 \,k_4 )
       + M_\pi^4   (  - 32 L_3^r L_5^r
\nn&&
 + 32 L_4^r L_5^r + 56 L_5^{r2} - 
         64 L_5^r L_6^r - 64 L_5^r L_8^r - 19/3 \,k_1 - 103/6 \,k_2 - 7/36 
         \,k_3 - 8 \,k_4
\nn&&
 - 55/12 \,k_5 + 4 \,k_7 + 2 \,k_8 )
       + M_\pi^2 M_K^2 q^2   ( 32 L_3^r L_5^r - 25/3 \,k_1 - 73/6 \,k_2
          - 205/36 \,k_3
\nn&&
 - 4 \,k_4 )
       + M_\pi^2 M_K^2   ( 24 L_3^r L_5^r + 64 L_4^r L_5^r - 32 L_5^{r2} - 128
          L_5^r L_6^r - 23/2 \,k_1 - 169/12 \,k_2
\nn&&
 - 181/36 \,k_3 + 2 \,k_4
          + 7/3 \,k_5 - 12 \,k_6 - 4 \,k_7 - 8 \,k_8 )
       + M_\pi^2 s_l q^2   ( 7/3 \,k_1 + 47/6 \,k_2 
\nn&&
 + 13/4 \,k_3 )
       + M_\pi^2 s_l   ( 8 L_3^r L_5^r + 8 L_5^r L_9^r + 13/6 \,k_1 + 35/4
          \,k_2 + 79/36 \,k_3 + 4/3 \,k_4
\nn&&
 - 3/4 \,k_5 - 11/8 \,k_9 )
       + M_\pi^2 \nu q^2   (  - 5/3 \,k_1 - 5/2 \,k_2 + 13/12 \,k_3
          )
       + M_\pi^2 \nu   ( 32 L_2^r L_5^r
\nn&&
 + 8 L_3^r L_5^r - 7/3 \,k_1 - 10 
         \,k_2 - 35/72 \,k_3 - 2/3 \,k_4 - 9/4 \,k_5 )
\nn&&
       + M_K^4  ( 8 L_3^r L_5^r + 8 L_5^{r2} - 29/12 \,k_1 - 13/8
          \,k_2 - 223/144 \,k_3 - 2 \,k_4 )
       + M_K^2 s_l   (  - 8 L_3^r L_5^r
\nn&&
 - 8 L_5^r L_9^r + 7/3 \,k_1 + 3 
         \,k_2 + 73/72 \,k_3 + 2/3 \,k_4 - 1/2 \,k_9 )
       + M_K^2 \nu   (  - 32 L_2^r L_5^r - 8 L_3^r L_5^r
\nn&&
 - 2/3 \,k_1 - 79
         /12 \,k_2 - 7/12 \,k_3 - 1/3 \,k_4 )
       + s_l^2   (  - 1/4 \,k_1 - 7/8 \,k_2 - 7/16 \,k_3 - 1/
         8 \,k_9 )
\nn&&
       + s_l \nu   ( 1/3 \,k_1 + 7/12 \,k_2 - 3/8 \,k_3 - 3/8 
         \,k_9 )
       + \nu^2   (  - 5/12 \,k_1 - 25/24 \,k_2 - 7/16 \,k_3 )~.
\nn&&
\eea

\noindent
\paragraph{5.}
We will now show some numerical results using the previous
formulas. It should be kept in mind that the final numbers are
quite sensitive to the input numbers and that there are several 
uncertainties:
\begin{enumerate}
\item The logarithms in the $k_i$ can have varying scales in them,
i.e. the scale $M$ in Eq. (\ref{eq:ki}) can be varied. Since all the
numerical examples refer to chiral $SU(3)$, the choice $M=M_K$ is the
most natural one. We will contrast the results for $M=M_K$ with those
for $M=\sqrt{M_K M_\pi}$. 
\item The final results are of course $\mu$ dependent. 
Both the values of $L_i^r(\mu)$ and the value of $L$ change as functions
of $\mu$.
\item The values of $L_i^r(\mu)$ that are used as input are in general
only determined via $p^4$ calculations.
\item The parts we included are dominant when $L$ is large. For the
$n=3$ calculations considered here, the logarithm typically is not as 
dominant as for $n=2$. Thus the remaining parts of the loop amplitudes
can be important.
\item The contributions from the $p^6$ Lagrangian are all set to zero.
\end{enumerate}

\begin{table}[t]
\begin{center}
\begin{tabular}{|c||c|c|c|c|}
\hline
change & set A & $\mu=0.9~$GeV & set B & $M=0.26~$GeV \\
\hline
\hline
$(F_\pi/F)^{(6)}$   & 0.013  & 0.029  &$-$0.050&$-$0.005\\
$(F_K/F_\pi)^{(6)}$ & 0.08   & 0.12   &0.06    &0.24    \\
\hline
$10^3\cdot f_{+1}$  & 3.6    &$-$4.8  &2.0     &$-$2.4  \\
$f_{+2}$[GeV$^{-2}$]&$-$0.28 &$-$0.24 &$-$0.25 &$-$0.26 \\
$f_{+3}$[GeV$^{-4}$]&0.58    &0.77    &0.55    &1.5     \\
$f_{02}$[GeV$^{-2}$]&0.30    &0.51    &0.18    &0.74    \\
$f_{03}$[GeV$^{-4}$]&0.26    &0.23    &0.41    &1.54    \\
\hline
$F_1$               &0.86    &1.13    &0.75    &2.5    \\
$F_2$               &0.38    &0.51    &0.17    &0.74    \\
$F_3$               &0.000   &$-$0.019&0.052   &0.14    \\
\hline
$G_1$               &$-$0.15 &$-$0.20 &$-$0.04 &0.18    \\
$G_2$               &0.006   &0.014   &0.035   &0.17    \\
$G_3$               &0.010   &0.012   &0.011   &0.031   \\
\hline
\end{tabular}
\caption{\label{tab:num} Numerical results for the partial $p^6$
corrections for several quantities. See text for definitions and input
parameters.}
\end{center}
\end{table}

For all these reasons the numbers quoted here should only be used
as an indication of the size of the $p^6$ corrections.
As input values we have used $F_\pi=92.4~$MeV, $M_K=495~$MeV,
$M_\pi=135~$MeV and 
$10^3\cdot L_{(4,5,6,8,9)}^r(0.77~\mbox{GeV}) =
(-0.3,1.4,-0.2,0.9,6.9)$.
We will vary $\mu$, the scale $M$ and use two sets
of $L_{(1,2,3)}^r$.
The first set corresponds to the unitarized fit of \cite{bcgke4},
$10^3\cdot L_{(1,2,3)}^r(0.77~\mbox{GeV}) =
(0.4,1.35,-3.5)$ (set A) whereas $10^3\cdot
L_{(1,2,3)}^r(0.77~\mbox{GeV}) = (0.6,1.5,-3.3)$ (set B) is from 
the one-loop fit of the same reference.
The values of $\mu=0.77~$GeV, $M=M_K=0.495~$GeV and set A are used unless
otherwise shown. We write for the $K_{l3}$ form factors 
\be
(f_+(t))^{(6)} = f_{+1}+f_{+2}t+f_{+3}t^2,\quad
(f_0(t))^{(6)} = f_{01}+f_{02}t+f_{03}t^2
\ee
and we evaluate the $K_{e4}$ form factors at $s_l=\nu=0$:
\be
F^{(6)}(q^2,0,0)= F_1 + F_2 q^2 +F_3 q^4,\quad
G^{(6)}(q^2,0,0)= G_1 + G_2 q^2 +G_3 q^4
\ee
The results are shown in Table \ref{tab:num}. Since by definition of
the $K_{l3}$ form factors $f_{01}=f_{+1}$ we do not display $f_{01}$.
The last column shows that $M=\sqrt{M_K M_\pi}=260~$MeV is probably
not a good choice for terms of the type
$\left(M_K^2\ln(M/\mu)\right)^2$ leading to unreasonably big numbers
in most cases. Not surprisingly, the numbers become even more 
unreasonable for $M=M_\pi$. 

The general size of the correction to $F_\pi/F$
is similar to the full result for the two-flavour case obtained
in \cite{bct98}. The correction to $F_K/F_\pi$ is rather large
and will, if the full $p^6$ calculation is of similar size, require
a revision of the value for $L_5^r$. 

As for the $K_{l3}$ form factors, the small correction $f_{+1}$ to 
$f_+(0)$ is an important result. The form factor at $t=0$ is 
used as input in the determination of the Cabibbo angle from $K_{l3}$ 
decays. It is therefore important to know that $f_+(0)$ is well
protected from $p^6$ corrections \cite{lr84}. The corrections to the
slopes $\lambda_+$, $\lambda_0$ ($f_{+2}$, $f_{02}$ in the table) have
about the same size but opposite signs. These corrections are within the
expectations for chiral $SU(3)$. However, the scalar slope might be more
sensitive to $p^6$ corrections because the $p^4$ prediction for
$\lambda_0$ is smaller than for $\lambda_+$ \cite{gl85b}. Finally, the
curvatures induced at $O(p^6)$ are completely negligible in the
physical region, in agreement with the observed Dalitz plot
distributions and with theoretical expectations \cite{gl85b}.

The size of the corrections to the $K_{e4}$ form factors
is such that a full $p^6$ calculation  
is desirable in order to refit $L_1$, $L_2$ and $L_3$. This is
especially true for the slope of the form factor $F$ ($F_2$ in the table)
where the $p^6$ correction is of the same size as the observed
slope \cite{rosselet}. Using the other inputs as above and fitting
$L_{(1,2,3)}$ to
$F(0,0,0) = 5.59\pm0.14$, $G(0,0,0) = 4.77\pm0.27$
and $(F(q^2,0,0)-F(0,0,0))/q^2|_{q^2=0.1}=(0.08\pm0.02)\,F(0,0,0)$
leads to $10^3\cdot L^r_{(1,2,3)}=(0.25\pm0.3,1.24\pm0.3,-4.75\pm1.9)$
to be compared with the one-loop $K_{e4}$-only fit of \cite{bcgke4} 
with the same inputs of $(0.65\pm0.3,1.63\pm0.28,-3.4\pm1.0)$.

\noindent
\paragraph{6.} We have determined the full double-pole divergence
structure of chiral perturbation theory at $O(p^6)$. We have
described the general method for extracting all contributions with
double chiral logs ($L^2$) as well as the full dependence
on $L_i^r\times L$ and $L_i^r\times L_j^r$. These partial $p^6$
corrections were then calculated for several quantities of physical 
interest. The corrections to $F_K/F_\pi$ and to the slopes of
the $K_{l3}$ form factors are of the size expected for chiral $SU(3)$.
On the other hand, we find small corrections to the $K_{l3}$ 
form factor $f_+$ at $t=0$, important for the
determination of $V_{us}$. The corrections for $K_{e4}$
decays are significant suggesting possible shifts
in the values of $L_{(1,2,3)}^r$.

\vspace*{0.5cm}

\noindent{\bf Acknowledgements}
We thank J.~Gasser for participation in the early stages of this
work and for continuous encouragement.


\end{document}